# Origin of large negative electrocaloric effect in antiferroelectric $PbZrO_3$

Pablo Vales-Castro[1*], Romain Faye[2], Miquel Vellvehi[3], Youri Nouchokgwe[2,4], Xavier Perpiñà[3], J.M.Caicedo[1], Xavier Jordà[3], Krystian Roleder[5], Dariusz Kajewski[5], Amador Perez-Tomas[1], Emmanuel Defay[2], Gustau Catalan[1,6*]

[1] Catalan Institute of Nanoscience and Nanotechnology (ICN2), Campus Universitat Autonoma de Barcelona, Bellaterra 08193, Spain; email: pablo.vales@icn2.cat

[2] Materials Research and Technology Department, Luxembourg Institute of Science and Technology, rue du Brill 41, Belvaux, L-4422, Luxembourg

[3] Institut de Microelectrònica de Barcelona (IMB-CNM,CSIC) Campus Universitat Autònoma de Barcelona (UAB), Cerdanyola del Vallès, 08193, Spain

[4] University of Luxembourg, 2 avenue de l'Université, L-4365 Esch-sur-Alzette, Luxembourg

[5] Institute of Physics, University of Silesia in Katowice, ul. Uniwersytecka 4, 40-00 Katowice, Poland

[6] Institut Català de Recerca i Estudis Avançats (ICREA), Barcelona 08010, Catalunya; email: gustau.catalan@icn2.cat

**Abstract**

We have studied the electrocaloric response of the archetypal antiferroelectric $PbZrO_3$ as a function of voltage and temperature in the vicinity of its antiferroelectric-paraelectric phase transition. Large electrocaloric effects of opposite signs, ranging from an electro-cooling of -3.5 K to an electro-heating of +5.5 K, were directly measured with an infrared camera. We show by calorimetric and electromechanical measurements that the large negative electrocaloric effect comes from an endothermic antiferroelectric-ferroelectric switching, in contrast to dipole destabilization of the antiparallel lattice, previously proposed as an explanation for the negative electrocaloric effect of antiferroelectrics.

Keywords: antiferroelectrics, electrocalorics, phase transitions, calorimetry, infrared thermometry

# I. Introduction

The electrocaloric effect (ECE) is the reversible temperature change (ΔT) of a material when a voltage step is applied or removed adiabatically [1]. It was first theorized in 1878 by William Thomson [2] as the inverse of the pyroelectric effect, but it took 50 years until the ECE was first observed in ferroelectric Rochelle Salt [3], and it was first quantitatively measured even later, by Hautzenlaub in 1943 [4]. Although initially it did not attract much attention because of the low temperature increments achieved, a large EC temperature change was calculated in 2006 for ferroelectric thin films [5], prompting a surge of interest in this effect. The ECE is attractive as a way to develop solid state cooling systems, and also because the theoretical efficiency goes up to 70%, much higher than thermoelectrics (10%) or even a conventional gas-cooling cycle (50%) [1]. Moreover, it has a great potential for scalability, useful to cool down advanced integrated circuits or complex systems in ever more powerful chips and heating-prone computers, including emerging wide-bandgap technologies (e.g. SiC, GaN or $Ga_2O_3$) that can operate at larger temperatures than the 175°C limit of silicon devices [6]. The scalability of the electrocaloric effect comes from the fact that the large electric fields required to produce large temperature changes can be achieved with modest voltages in thin films, thanks to their reduced thickness and increased breakdown strength [5].

Antiferroelectrics (AFE) are materials with antiparallel sub-lattices of electric dipoles that can be switched under electric field into a polar state. Their electrocaloric properties have been less investigated than those of ferroelectrics, but their study has increased since 2011 with the discovery of the anomalous electrocaloric effect [7] (also called negative electrocaloric effect), whereby applying a voltage causes a decrease in temperature (ΔT< 0), rather than an increase. This "electrocooling" is contrary to the normal (or positive) ECE displayed by conventional ferroelectrics, which increase their temperature (ΔT > 0) when the field is applied. This electrocaloric temperature change ΔT can be extracted from Maxwell relations as defined by equation (1) (see Supplemental Material [8] and references [9–13]) . The interest in this anomalous effect was further enhanced by the report of indirectly-measured "giant" negative electrocaloric effects in antiferroelectric thin films [10,14].

$$\Delta T = -\int_{E_1}^{E_2} \frac{T}{C_E(T,E)} \left(\frac{\partial D}{\partial T}\right)_E d\mathrm{E} \qquad (1)$$

## II. Microscopic electrocaloric models

It may seem surprising that a material can get colder when energy (voltage, in this case) is added to it. While macroscopically the negative electrocaloric effect occurs because $\left(\frac{\partial D}{\partial T}\right)_E > 0$ (equation 1), the microscopic mechanism that enables such behaviour in antiferroelectrics is still being debated, and there are at least two possible explanations put forward in the community. On the one hand, when an electric field is adiabatically applied to an antiferroelectric, it destabilises the dipole sub-lattice that is antiparallel to the applied field, as explained by Geng et al. [10,14] thus increasing its dipolar entropy ($S_{dip}$) and thereby reducing its temperature (by decreasing the phononic contribution $S_{ph}$) to satisfy equation (1).

$$\Delta S = \Delta S_{dip} + \Delta S_{ph} = 0 \quad (2)$$

This is opposite to paraelectrics and ferroelectrics, where electric fields increase dipole alignment and thus reduce dipolar entropy (parenthetically, negative ECE can also appear in ferroelectrics when the polarization is antiparallel or not collinear with the applied electric field [15]).

On the other hand, a negative electrocaloric effect can also appear in any material that undergoes a field-induced first order phase transition between two phases, as already reported for some ferroelectrics [15–17] and also for antiferroelectric $PbZrO_3$ [18]. This can take place as long as the field-induced phase transition is defined by a positive latent heat $L_H$ (defined by $\Delta S_{LH}$ ) and as long as this is larger than the entropy variation of the smooth and continuous change of P through the transition ($\Delta S_{\tilde{P}}$), according to equation (2) [19]. Note that equation (2) is just a generalization of equation (1) when latent heat has a role in the transition.

$$\Delta S = \Delta S_{dip} + \Delta S_{ph} = (\Delta S_{\tilde{P}} + \Delta S_{LH}) + \Delta S_{ph} = 0 \quad (3)$$

In this framework, the negative ECE in the prototypical antiferroelectric $PbZrO_3$ would be best described not so much as a destabilization of a polar sub-lattice, but as a field-induced endothermic phase transition.

These two models predict rather different functionalities. In the sublattice de-stabilization ("dipole canting") model, once the antiparallel sub-lattice is switched, its entropy should decrease again as the dipoles have been re-aligned parallel to the field. The cooling achieved during the sub-coercive part, i.e., below the antiferroelectric-ferroelectric transition field ($E_{AFE\text{-}FE}$) should therefore turn to heating upon switching ($E > E_{AFE\text{-}FE}$), and the net thermal balance after saturation would tend to be neutral or

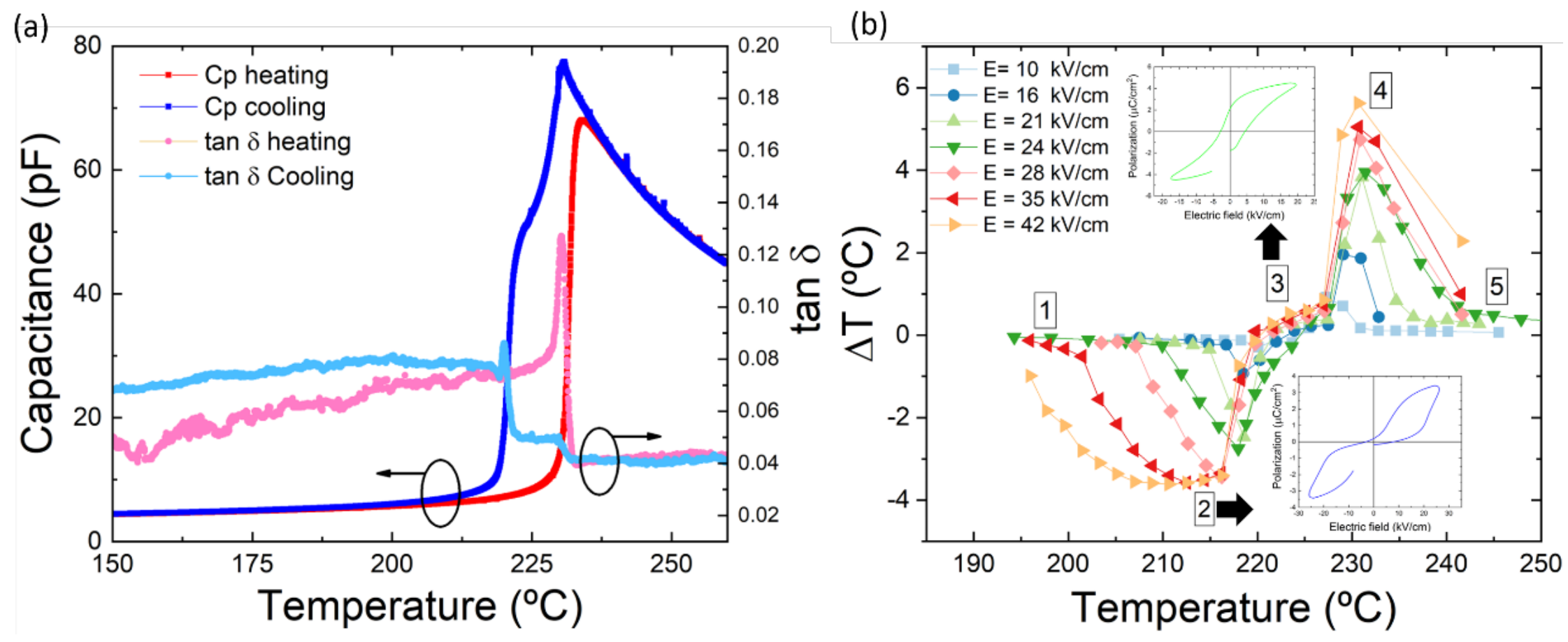

*Figure 1 (a) Capacitance and losses of $PbZrO_3$ ceramics and (b) electrocaloric temperature change versus temperature of $PbZrO_3$, characterized with an IR camera. Insets represent the polarization versus electric field hysteresis loops.*

positive [10]. By contrast, if the negative electrocaloric effect relies on latent heat of transformation, cooling will be expected to increase, rather than decrease, when there is antiferroelectric switching until it saturates. Put another way: while the dipole-canting model relies on sub-switching fields ($E < E_{AFE\text{-}FE}$), the phase-change model requires overcoming the phase transition point ($E > E_{AFE\text{-}FE}$). Clarifying which is the dominant contribution to the negative ECE of antiferroelectrics is therefore not only a fundamental science question; it is also essential for maximizing the response of electrocaloric devices as a function of voltage. Determining which of the two models (dipole canting [14] or latent heat of transformation [18]) dictates the large negative electrocaloric response of the archetypal antiferroelectric ($PbZrO_3$) is the aim of this investigation.

## III. Results

### A. Dielectric and electrocaloric measurements

Antiferroelectric $PbZrO_3$ ceramics were fabricated as reported in [20]. Their capacitance and losses as a function of temperature are shown in figure 1-a. On heating, there is a single peak at the Curie temperature ($T_c$), signalling the transition from the antiferroelectric (orthorhombic, Pbam symmetry [21]) phase to the paraelectric (simple cubic, Pm3m symmetry [21]) one. On cooling, there is an additional shoulder at lower temperatures suggesting the existence of a ferroelectric (rhombohedral, R3c or R3m symmetry [22]) stable intermediate phase. For the sake of simplicity, we concentrate the discussion on the electrocaloric response on heating, without loss of generality; the full set of results for heating and cooling are provided in Supplemental Materials [8] (Section S1)

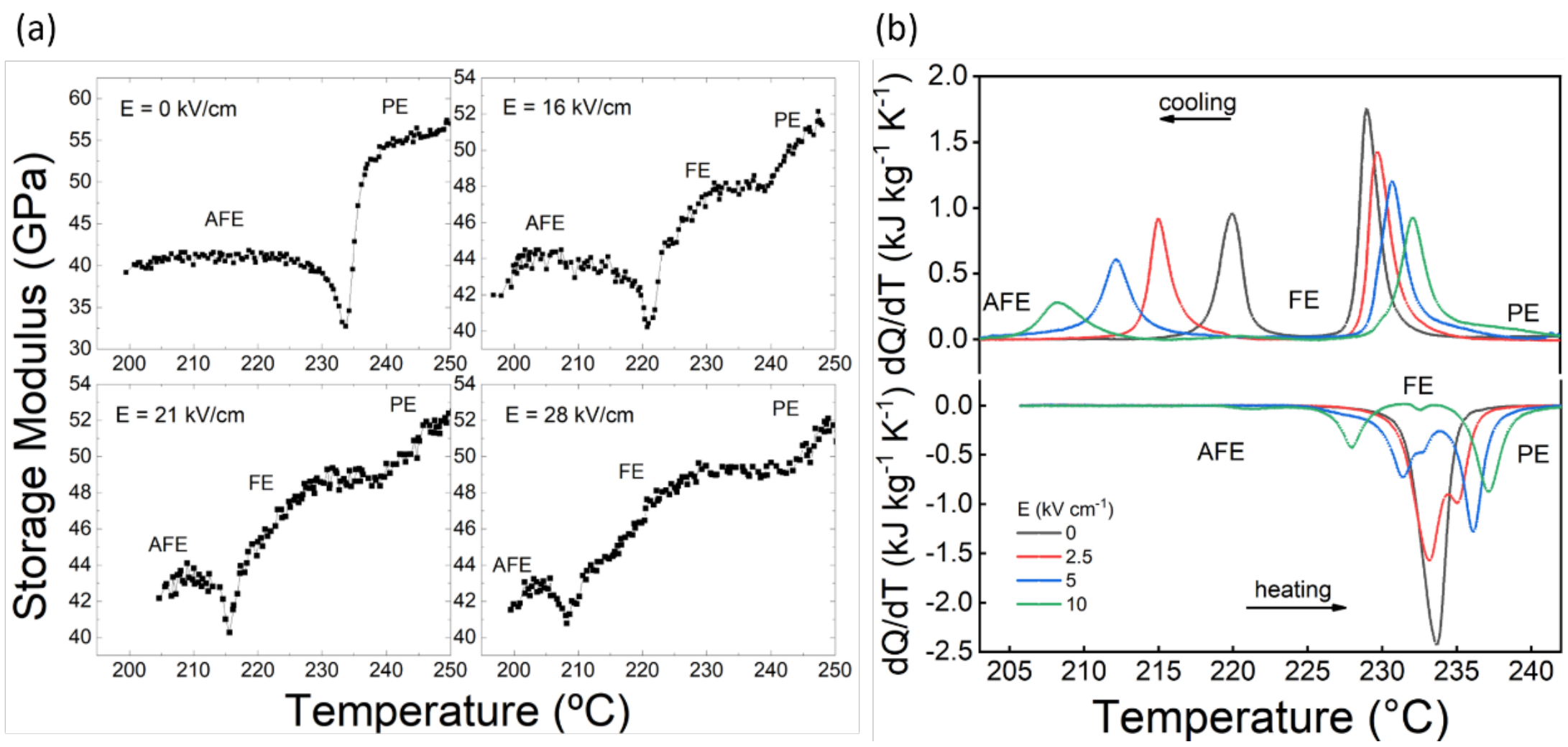

*Figure 2 (a) Storage modulus of a PZO ceramic vs temperature obtained with Dynamic Mechanical Analysis (DMA). The discontinuities represent switching between different phases of PZO; (b) Differential Scanning Calorimetry (DSC) measurements of a PbZrO3 ceramic at different electric fields for heating and cooling*

The electrocaloric response as a function of temperature and field is shown in figure 1-b, together with the relevant polarization versus field loops at different regions. We have labelled the ranges where qualitatively and quantitatively different behaviours are displayed. At low temperatures and/or with low voltages (range 1), there is only a small negative ECE ($\Delta T \leq -0.6$ K). Above a temperature-dependent critical field, there is a jump in the negative response (range 2), reaching a maximum temperature change of $\Delta T = -3.6$ K for fields $\geq 35 kVcm^{-1}$. The maximum negative electrocaloric strength is very high, peaking at $(|\Delta T||\Delta E|^{-1})_{negECE} = 0.12$ K cm $kV^{-1}$. At higher temperatures, the effect abruptly changes from large negative to almost zero or weakly positive in the temperature range between ~220 °C and ~227 °C (range 3). Then there is another sharp increase, whereby the ECE abruptly rises to a positive peak (range 4) before dropping again to weakly positive (almost zero) above $T_C$ (range 5). The maximum positive ECE in range 4 was $\Delta T = +5.6$ K, and the maximum electrocaloric strength $(|\Delta T||\Delta E|^{-1})_{posECE} = 0.18$ K cm $kV^{-1}$. The overall response in the whole temperature span (i.e regions 1 through 5) is qualitatively similar to that predicted for Ba-doped $PbZrO_3$ ceramics [23], for which doping stabilizes the FE intermediate phase over a wider temperature range.

The electrocaloric maxima (both negative and positive) are listed in Table S1 (Supplemental Materials [8] and references [9,18,23–41]) and are compared to other directly measured values reported in the literature. The maximum negative electrocaloric effect in $PbZrO_3$ is significantly higher than any result previously reported by direct measurements. (Parenthetically, the positive electrocaloric peak is also among the highest, comparable to $PbSc_{0.5}Ta_{0.5}O_3$ multilayer capacitors [40]). This high electrocaloric performance of $PbZrO_3$ is intrinsic and not achieved through compositional fine-tuning: $PbZrO_3$ is a pure compound and the antiferroelectric archetype.

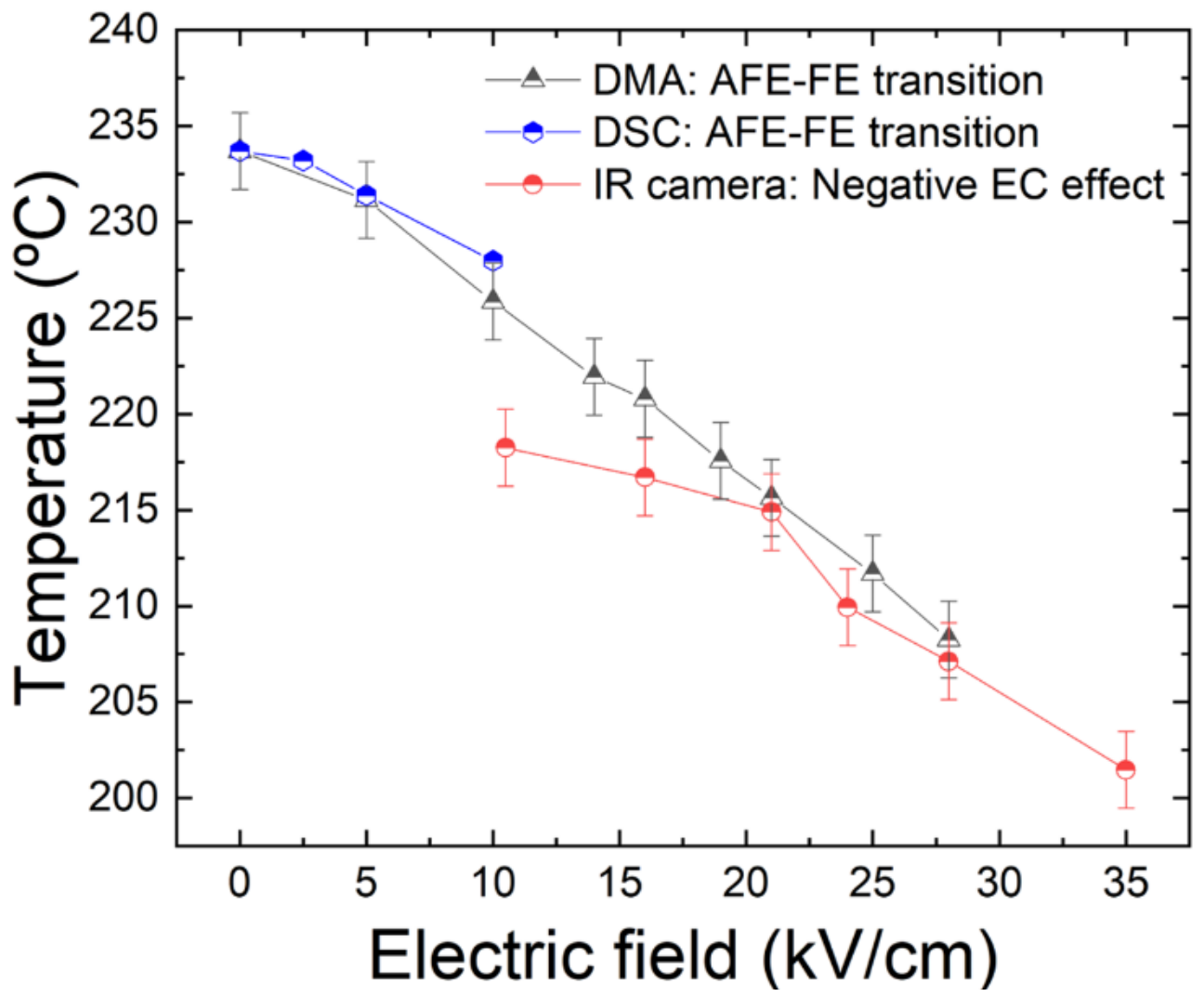

*Figure 3 Critical temperature of the AFE-FE phase transition as a function of field, measured by DMA (black) and DSC (blue), and onset temperature of giant negative electrocaloric effect measured by the infrared camera. The three independent measurements coincide.*

### B. Mechanical analysis and calorimetry

In order to determine the apparent link between electrocaloric response and field-induced phase transition, we have performed two additional experiments: dynamic mechanical analysis (DMA) as a function of temperature and electric field (figure 2-a) and differential scanning calorimetry (DSC) also as a function of electric field (figure 2-b).

Dynamic mechanical analysis measures the mechanical storage modulus and is very sensitive to structural phase transitions [42]. Under zero field, $PbZrO_3$ on heating shows only one sharp minimum at the antiferroelectric-paraelectric (AFE-PE) phase transition, as expected, and consistent also with the dielectric constant measurements (figure 1-a). With voltage, the mechanical singularity splits into two: one that shifts to higher temperatures (consistent with a FE-PE transition) while the other moves to lower temperatures with increasing field (consistent with an AFE-FE transition).

It is interesting to notice that the field-induced polar phase can indeed be stable (ferroelectric-like), as shown by the ferroelectric hysteresis loops (inset of figure 1-b). Even in the absence of external field, this ferroelectric phase can appear on cooling [43–47], and is responsible for the second anomaly of the dielectric constant (figure 1-a); on heating, however, the ferroelectric phase only appears when a high enough energetic external stimulus, such as a voltage, is applied to the system. Therefore, if the polar phase is unstable, removal of the field returns the material to its antipolar phase, yielding the typical antiferroelectric double-hysteresis loop. If it is stable, however, the material stays “locked” into a ferroelectric state even after the field is removed, so subsequent voltage pulses do not modify the polar state, yielding a standard FE hysteresis loop. This behaviour translates into different electrocaloric responses.

Differential Scanning Calorimetry (DSC) measurements in figure 2-b show the heat flow dQ/dT of $PbZrO_3$ bulk ceramic at four different electric fields as a function of temperature, both on heating and on cooling. A single endothermic peak on heating is observed at 0 kV $cm^{-1}$ (black curve in figure 2-b). This peak corresponds to an endothermic AFE-PE first-order transition (latent heat), and is consistent both with the dielectric (figure 1-a) and electromechanical results (figure 2-a). Like the DMA, the DSC also shows this splitting into two distinct peaks with the increasing electric field. The lower-temperature one, which corresponds to the AFE-FE phase transition, shifts towards ever-lower temperatures with increasing field. The second peak (FE-PE) moves towards high temperatures. Thereby, applying an electric field on our bulk $PbZrO_3$ stabilizes the ferroelectric phase (Supplemental Materials [8], figure S10). This behaviour is analogous to that of DMA (Figure 2-a).

## IV. Discussion

Based on the presented results, we can now draw a conclusion on the origin of the large electrocaloric effect. Qualitatively, we observe that the transitions between weak (region 1) and large responses (region 2) are abrupt (figure 1-b), and the AFE and FE hysteresis loops (inset in figure 1-b) exceed the phase transition field $E_{AFE-FE}$ and display saturation. These observations, together with DSC data thus provide definitive evidence for the link between “giant” electrocaloric effects (both negative

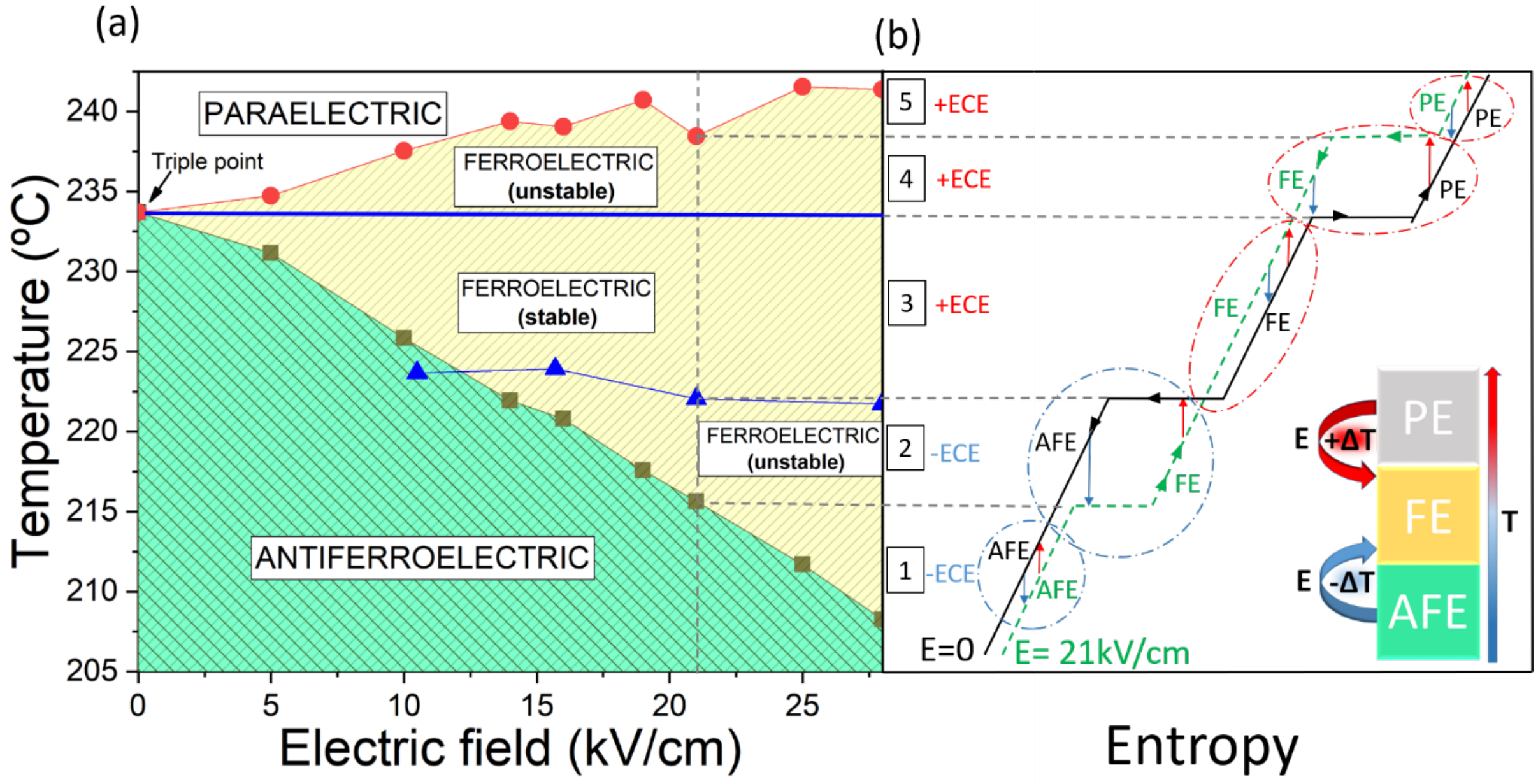

*Figure 4 (a) Tentative phase diagram for $PbZrO_3$ as a function of electric field. AFE-FE and FE-PE transitions are plotted with electromechanical results (DMA) while the Ferroelectric/Field-induced ferroelectric transition is plotted with electrocaloric data (IR camera). (b) Schematic T-S diagram of $PbZrO_3$ showing all phase transitions for E = 0 and E = 21 kV/cm. The arrows represent the adiabatic (ΔS = 0) electrocaloric ΔT when an electric field E is applied or removed. All the different electrocaloric scenarios are labelled as in Figure (1-b). The inset in (b) represents the connection between the phase transitions in $PbZrO_3$ and its positive and negative electrocaloric responses.*

and positive) and field-induced phase transitions: as displayed by the DSC data (Figure 2-b) the AFE-FE peak corresponds to an endothermic transition, which yields cooling upon field application, i.e. a negative electrocaloric effect (region 2 in Figure 1-b). In fact, the large negative electrocaloric effect in region 2 starts precisely at the phase transition temperatures given by DMA and DSC, as shown in Figure 3. Conversely, the PE-FE peak is exothermic and thus linked to a positive electrocaloric effect (region 4 in Figure 1-a). All evidence thus indicates that, even if there is some dipole canting contribution [8] (range 1 in figure 1-b) above the AFE-FE transition point ($E > E_{AFE-FE}$), its role is negligible compared to that of the latent heat of transformation. In addition, our maximum negative ΔT matches atomistic calculations for the antiferroelectric-ferroelectric phase boundary [48], also consistent with a first-order phase transition origin.

Based on the different measurements, it is possible to draw a temperature-field phase diagram for PbZrO3 (figure 4-a) including the three phases: antiferroelectric, ferroelectric, and paraelectric. Based on this phase diagram, we have also generated a thermodynamic scheme showing how the electrocaloric effect works in the different regimes (figure 4-b and figure S7-b); this scheme should be valid for any antiferroelectric with an endothermic AFE-FE phase transition:

(1) Region 1 corresponds to a dipole canting response where no phase transition takes place. Thus, we have a reversible system where the negative ΔT achieved when the voltage is turned on ($V_{ON}$) reverses into positive ΔT of similar magnitude when it is turned off ($V_{OFF}$).
(2) Region 2 is linked to the endothermic AFE-FE phase transition (figure 2-b). Notice that the expected field-induced temperature changes are not symmetric, and this has been experimentally confirmed (Supplemental Materials [8], figure S7-a and figure S9-b).
(3) In region 3 the FE phase is stable, and thus a regular positive ECE takes place, with a symmetric $V_{ON}$/ $V_{OFF}$ response (figure S1-S6 and figure S8-e).
(4) Region 4 corresponds to the PE-FE phase transition upon field application, that is, the large positive ECE response typical of ferroelectrics close to $T_C$.
(5) Region 5 yields a low positive ECE following the same standard electrocaloric mechanism as in Region 3: slight increase in dipole alignment yields slight changes in temperature.

Therefore, we have answered the basic question about the origin of the so-called "giant" negative ECE in antiferroelectric $PbZrO_3$: this large response (ΔT = -3.5K) is a latent-heat mediated ECE coming from the first-order AFE-FE phase transition. In contrast, smaller negative ECE responses (ΔT ≲ -0.6 K) have its origin in the dipolar de-stabilization (canting) model.

Our results for pure $PbZrO_3$ are opposite to those measured in La-doped $Pb(Zr,Ti)O_3$ (PLZT) [49,50] or $Pb_{0.99}Nb_{0.02}[(Zr_{0.58}Sn_{0.43})_{0.92}Ti_{0.08}]_{0.98}O_3$ (PNZST) antiferroelectrics [41] which display a positive electrocaloric effect when the AFE-FE transition takes place ($E > E_{AFE\text{-}FE}$). The difference seems to be correlated with the doping-induced change in the position of the FE phase relative to the AFE one in the material's phase diagram [51], where a negative ECE by indirect methods was achieved in an AFE-FE phase sequence in PNZST 13/2/2. While the low-temperature phase (i.e., the ground state) is antiferroelectric for $PbZrO_3$ and Ba-doped $PbZrO_3$ [23], it is ferroelectric for PLZT and PNZST. Hence, the polar phase has higher entropy and therefore the field-induced transition to the polar phase is endothermic for $PbZrO_3$, while the opposite is true for PLZT and PNZST. In fact, first principle calculations also on the archetype antiferroelectric $PbZrO_3$ [52], but in this case with a FE-AFE-PE phase sequence i.e FE phase as the ground state, shows that the AFE-FE phase transition yields a positive electrocaloric effect, contrary to our negative response.

Our direct measurements also show a very large positive electro-caloric peak ($\Delta T = +5.6$ K) in $PbZrO_3$, linked to the transition between the paraelectric phase and the intermediate ferroelectric phase. The link between the giant negative response and antiferroelectric switching not only affects the field-range in which the response is maximized, but also has consequences for the useful temperature range. As long as the external field is bigger than the antiferroelectric-ferroelectric phase transition field $E_{AFE-FE}$ and smaller than the breakdown field, there will be switching and hence a large negative ECE. This is in contrast to the equivalent positive ECE peak of ferroelectrics, which is tied to their field-induced paraelectric-ferroelectric transition and thus anchored to their Curie temperature. In our $PbZrO_3$ ceramics, the maximum applied field was $E = 42$ kV $cm^{-1}$ (Figure 1-b), sufficient to cause switching and concomitant large ECE over a range of ~20 K below the Curie temperature, but in thin films it is possible to achieve antiferroelectric switching even at room temperature [53]. Since the absolute value of the ECE is directly proportional to the absolute temperature of the sample (equation 1), a reduction in the ECE is still expected at room temperature, proportional to the ratio between room temperature and Curie temperature, which is about 60% for the specific case of $PbZrO_3$. Nevertheless, this is still a substantial negative ECE that can in theory be retained over a temperature range two hundred degrees below the Curie point. The link to antiferroelectric switching thus potentially gives the negative electrocaloric effect a practical advantage over the positive effect of ferroelectrics in terms of wide temperature range of application.

## V. Summary

The present work shows that the large electrocaloric effect of antiferroelectric $PbZrO_3$ does not arise from a continuous destabilization of the antiparallel sublattice, but from the field-induced first-order endothermal transition between the antiferroelectric and ferroelecric phases. This has important implications for the functional response of the material: it means that the effect does not disappear above coercive fields, it links the dynamics of the electrocaloric response to the switching dynamics of the antiferroelectrics, and it implies that large negative electrocaloric effects can in principle be achieved over the entire field-temperature range of the AFE-FE transition.

## VI. Materials and experimental methods

### Fabrication and sample preparation

We have measured the electrocaloric effect of ceramics of the perovskite antiferroelectric archetype, $PbZrO_3$. The samples were made as described in [20] and disk-polished down to thicknesses between 100 and 150 microns with a Multiprep polishing system. Platinum electrodes where deposited by electron beam evaporation and platinum wires bonded with curated silver paste.

### Dielectric measurements

The capacitance and losses were measured (Agilent Precision LCR Meter, Model E-4980A) as a function of temperature to establish their quality (low losses) and pin-point the antiferroelectric-paraelectric phase transition. Polar hysteresis loops were measured with a Radiant LC meter at 1 kHz in order to establish the antiferroelectric/ferroelectric/dielectric nature of the different phases. The samples' temperature was controlled in a Linkam system in vacuum to increase the air breakdown field.

### Dynamic Mechanical Analysis (DMA)

Mechanical properties were done using a Perkin Elmer Dynamic Mechanical Analyzer (DMA) model 8, equipped with an in-situ oven in a three-point bending geometry. Samples were connected to an external high-voltage source while mechanically stressed between 1-3 Hz. To improve mechanical robustness, the samples used for mechanical analysis were thicker than those used for electrocaloric measurements, between 250 and 350 μm in thickness. The electric field was applied continuously to the sample as it heated up and cooled down. The error in absolute temperature of the samples (discrepancies between the thermocouples used in the different experimental setups) is ≤2 K.

### Infrared electrocaloric characterization

The electrocaloric performance was measured by infrared (IR) thermometry, using two different infrared cameras to ratify the robustness of the result. The cameras were a FLIR x6580sc and a FLIR SC5500 with acquisition speeds (frames per second, fps) of 130 and 376 fps, respectively, and a field of view of 3.2 mm x 2.55 mm. Prior to the IR characterization, the samples were covered with an emissivity-calibrated black paint. The temperature was controlled with a Linkam system. The electrocaloric effect was induced by voltage delivered by a Keithley High Voltage Sourcemeter 2410, with source current capped at 0.1 mA. The measuring process is based on the dynamics of a Brayton cycle: applying a voltage step adiabatically, acquire the response and let the sample thermalize before adiabatically removing the field. The relative temperature changes acquired with the IR camera were measured with an accuracy of 0.1 $^{o}$C.

**Differential Scanning Calorimetry (DSC)**

Using a commercial Differential Scanning Calorimeter (DSC), NEZTSCH, heat flux *dQ/dt* measurements at zero field and under three electric fields (2.5, 5, 10 kV cm$^{-1}$) were done on 6.00 mg bulk ceramic $PbZrO_3$ at a heating rate of 10 K min$^{-1}$. The electric field was maintained fixed in the material during the measurement**.** From heat flux measurements and after removal of the baseline we compute the heat flow measurements dQ/dT shown in figure 2-b.

**ACKNOWLEDGEMENTS**

We acknowledge financial support to ICN2, which is funded by the CERCA programme/ Generalitat de Catalunya and by the Severo Ochoa programme of the Spanish Ministry of Economy, Industry and Competitiveness (MINECO, grant SEV-2017-0706). We also acknowledge support to Plan Nacional (MINECO Grant MAT2016-77100-C2-1-P) and MINECO Grant BES-2016-077392. R. Faye, Y. Nouchokgwe and E. Defay thank the Luxembourg National Research Fund (FNR) for funding part of this research through the projects CAMELHEAT/C17/MS/11703691/Defay and COFERMAT/P12/4853155/Kreisel. This work was also supported in part by the Spanish Ministry of Science, Innovation and Universities under the HIPERCELLS project (RTI2018-098392-B-I00), the Regional Government of the Generalitat de Catalunya under Grant 2017 SGR 1384, and the Consejo Superior de Investigaciones Científicas through internal research programs with identification numbers 201850I063 and 201950E036. This work was also supported by the National Science Centre, Poland, within the project 2016/21/B/ST3/02242.

# SUPPLEMENTAL MATERIALS

## SECTION S1: COMPLETE ELECTROCALORIC DATA OF $PbZrO_3$

In the following sets of images (Figure S1-S6) we will show the complete results of the infrared characterization of the $PbZrO_3$ ceramic, measured following a Brayton cycle as explained in the Methods section. Specifically, for every set of images, (a) will display the $V_{ON}$ and $V_{OFF}$ electrocaloric response on heating while (b) will display the $V_{ON}$ and $V_{OFF}$ electrocaloric response on cooling as a function of the sample's temperature.

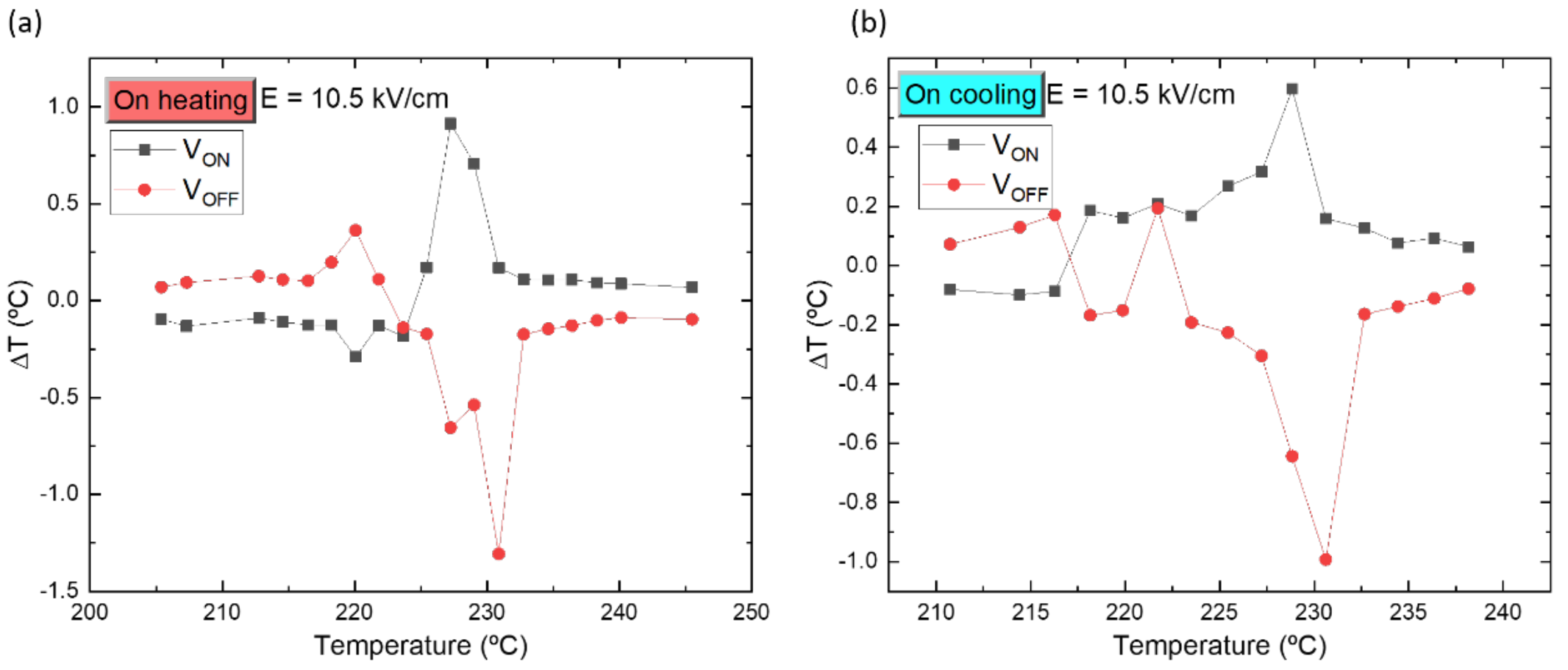

*Figure S1 Direct measurements of the electrocaloric effect in ceramic $PbZrO_3$ for an electric field E = 10.5 kV/cm. (a) ECE response in ON/OFF on heating and (b) ECE response in ON/OFF on cooling.*

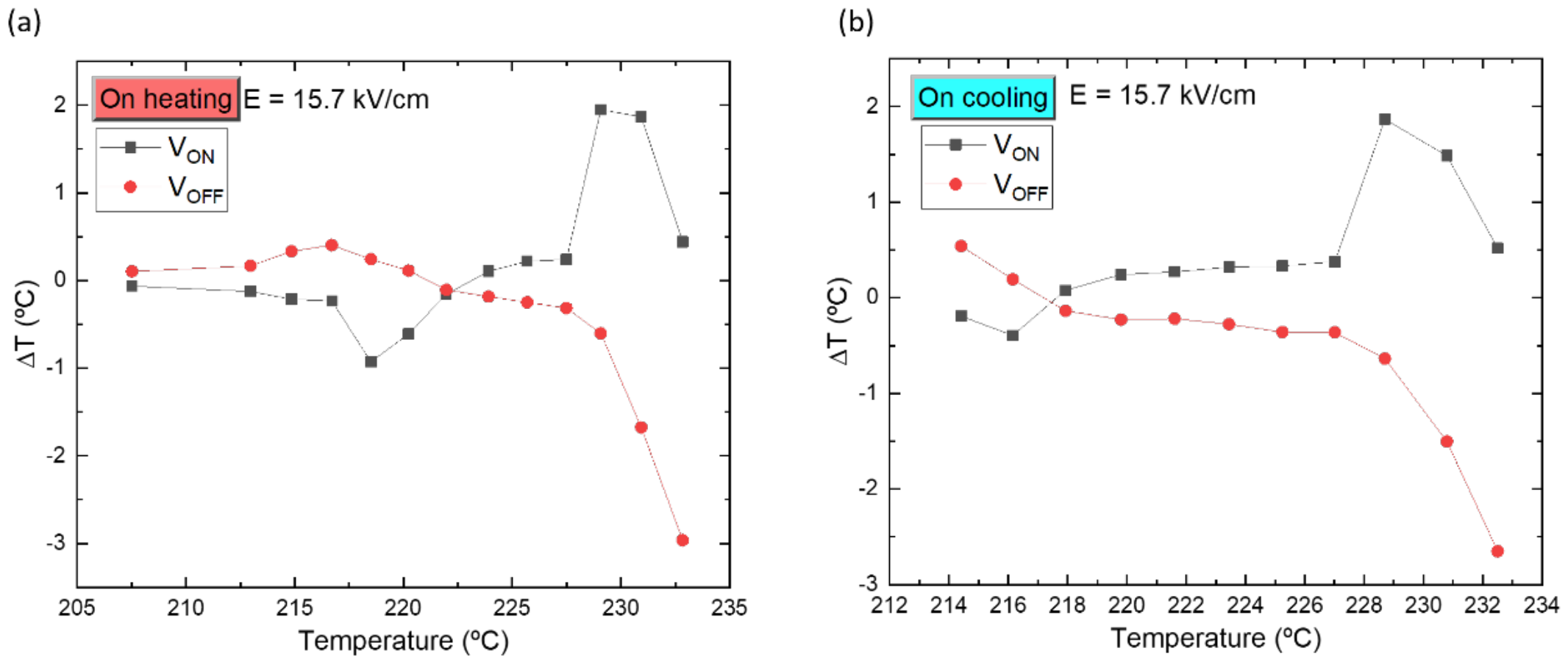

*Figure S2 Direct measurements of the electrocaloric effect in ceramic $PbZrO_3$ for an electric field E = 15.7 kV/cm. (a) ECE response in ON/OFF on heating and (b) ECE response in ON/OFF on cooling.*

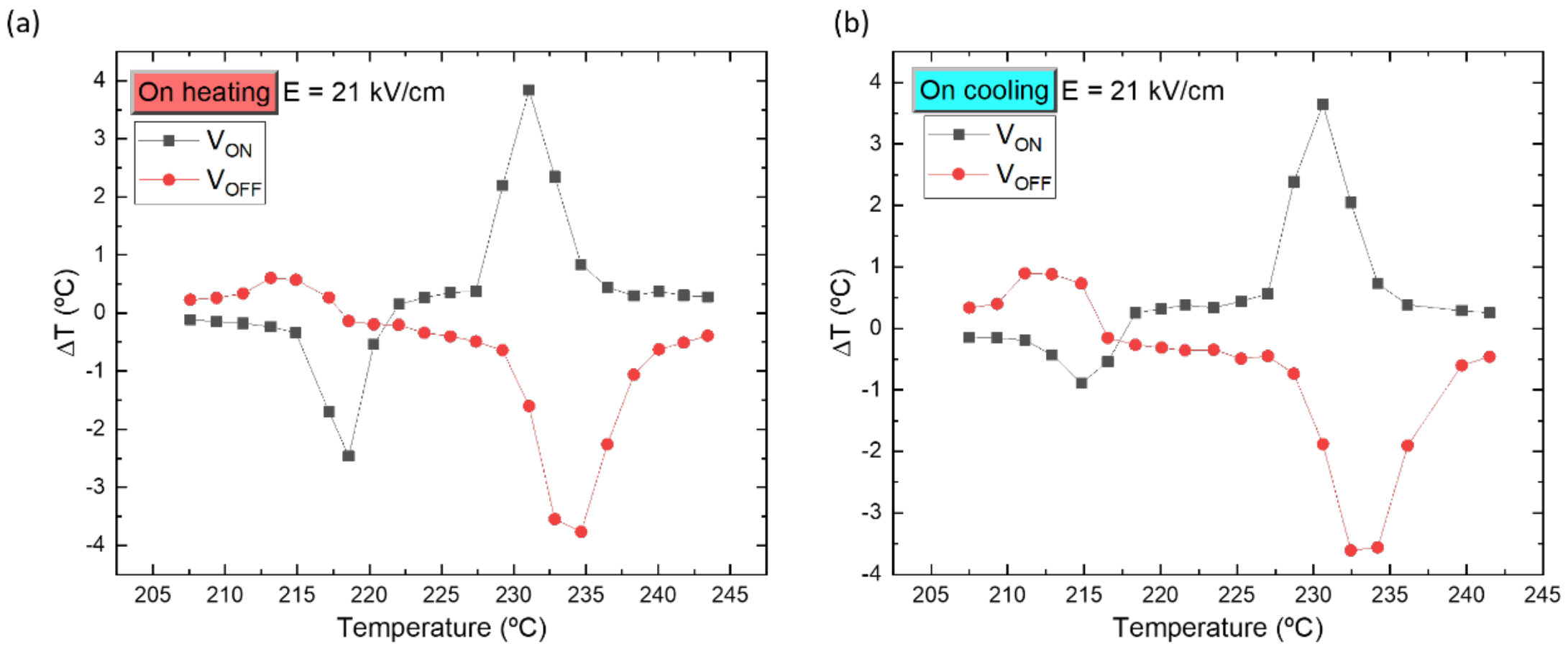

*Figure S3 Direct measurements of the electrocaloric effect in ceramic $PbZrO_3$ for an electric field E = 21 kV/cm. (a) ECE response in ON/OFF on heating and (b) ECE response in ON/OFF on cooling.*

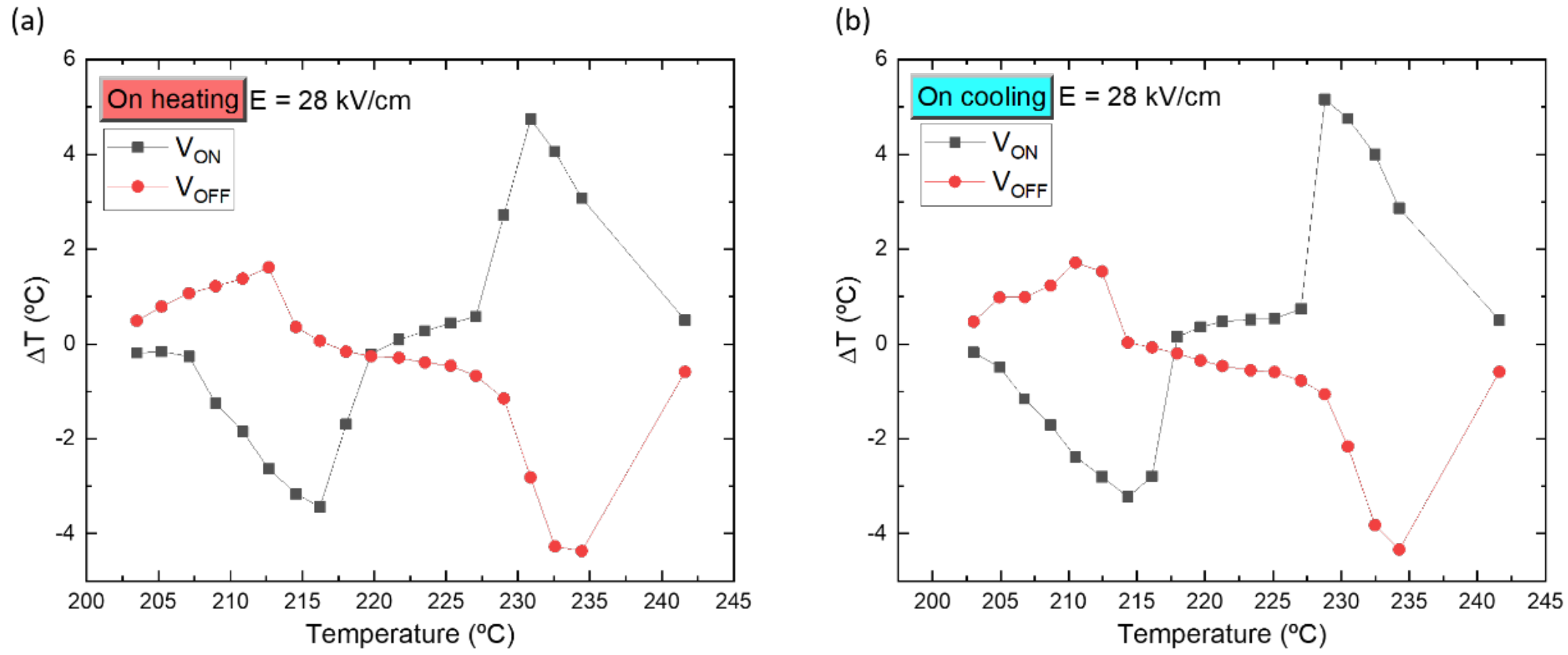

*Figure S4 Direct measurements of the electrocaloric effect in ceramic $PbZrO_3$ for an electric field E = 28 kV/cm. (a) ECE response in ON/OFF on heating and (b) ECE response in ON/OFF on cooling.*

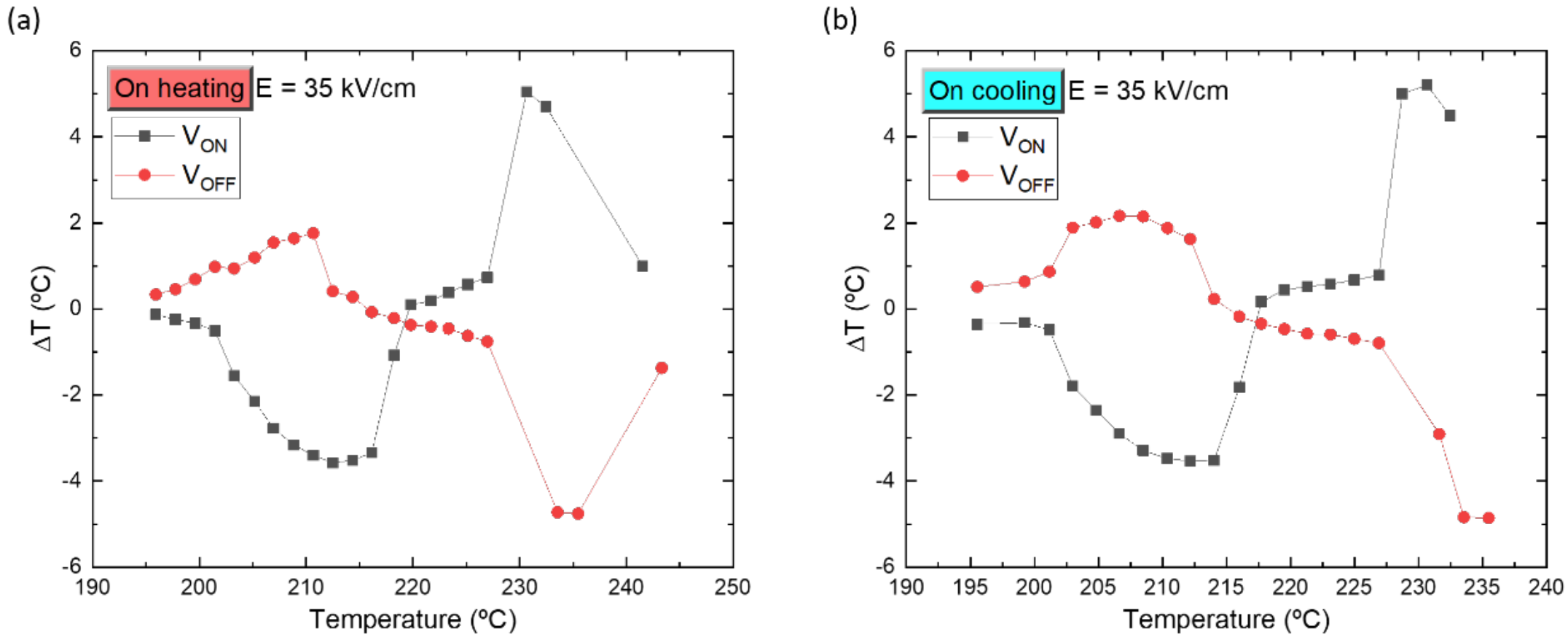

*Figure S5 Direct measurements of the electrocaloric effect in ceramic $PbZrO_3$ for an electric field E = 35 kV/cm. (a) ECE response in ON/OFF on heating and (b) ECE response in ON/OFF on cooling.*

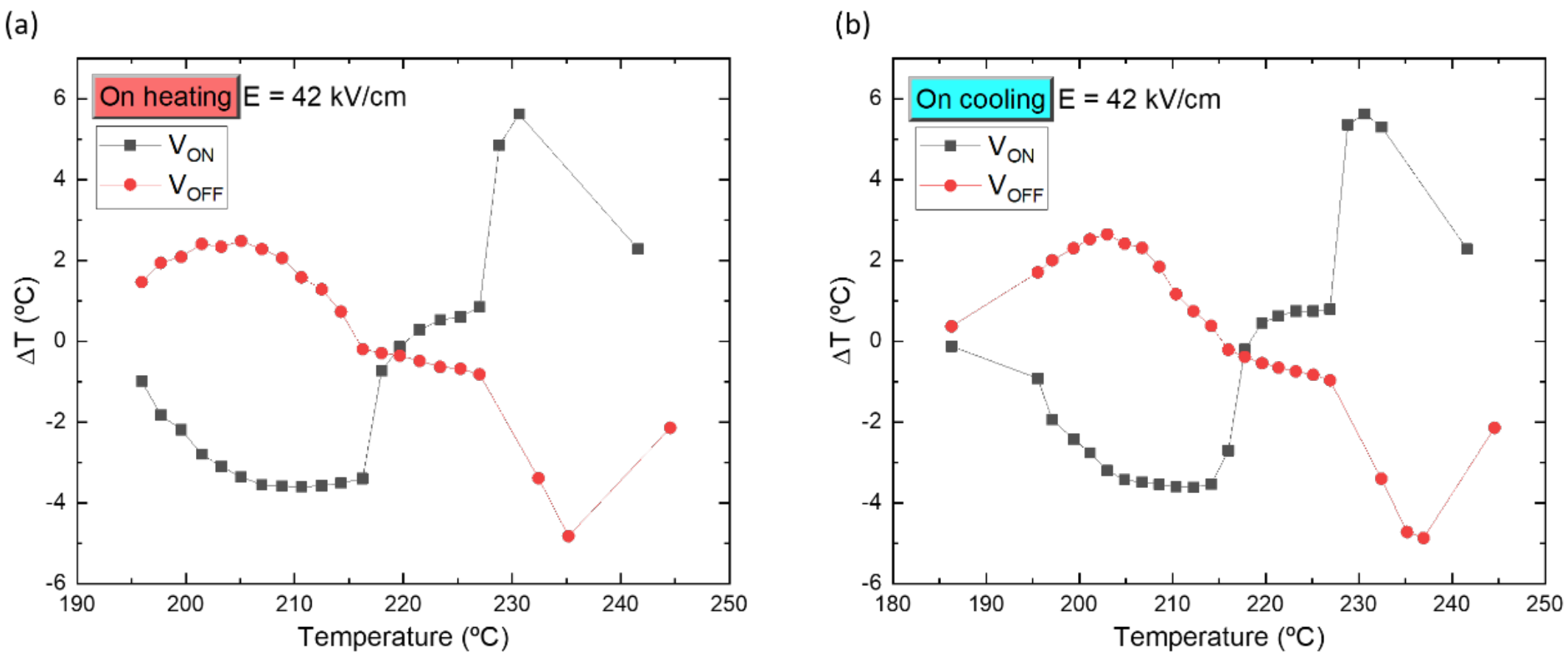

*Figure S6 Direct measurements of the electrocaloric effect in ceramic $PbZrO_3$ for an electric field E = 42 kV/cm. (a) ECE response in ON/OFF on heating and (b) ECE response in ON/OFF on cooling.*

Additionally, in table S1 we show our electrocaloric results compared to other directly measured values reported in the literature. We make a comparison only with direct methods because these are the most reliable in terms of absolute values of ΔT, but they are also the most scarce. In fact, most investigations in the literature tend to focus on thin films, where direct electrocaloric measurements are difficult, as they thermalize before their temperature change can be registered. Therefore, experiments are based on indirect measurements of dD/dT (where D is the dielectric displacement field –in effect, the field-induced polarization for non-ultrathin films) which are then translated into theoretical temperature changes via eq. (1) [1]:

$$dT = -\frac{T}{C_E}\left(\frac{\partial D_i}{\partial T}\right)_E d\mathrm{E} \qquad (1)$$

However, as previously stated, this indirect method has shortcomings that can modify the calculated ΔT with respect to the real value: (i) it does not take into account the energy dissipated by leakage currents, which can be high under high voltages. (ii) it assumes thermal equilibrium, whereas real experiments are time-dependent, particularly if switching is involved, where time evolution of the heat capacity and polarization may differ. (iii) The assumption of constant specific heat C as a function of T and E is problematic in systems that undergo first order phase transitions [2], as latent heat is released [3,4].

When the field-dependence of the specific heat is incorporated into eq. 1, a good agreement has been achieved for ferroelectrics [5]. However, the experimental characterization of the specific heat as a function of temperature and field, C(T, E), is not trivial, especially in antiferroelectrics where the polar phase can only be characterized under high electric fields. If C(E,T) is not known, equation (1) cannot be used accurately and the ECE should be measured by a direct method –which is thankfully possible in ceramics.

| Material | T (ºC) | ΔT (K) | ΔE ($kVcm^{-1}$) | ΔT $ΔE^{-1}$ ($K\ cm\ kV^{-1}$) | Ref. |
|---|---|---|---|---|---|
| $PZO^c$ | 97 | -1.5 | 80 | -0.0189 | [6] |
| $PNZST^c$ | 127 | -0.3 | 20 | -0.015 | [7] |
| Ba-doped $PZO^c$ | 107 | -0.6 | 40 | -0.015 | [8] |
| (001)-PMN-$30PT^s$ | ~80 | -0.13 | 2.5 | -0.052 | [9] |
| (001)-PMN-$30PT^s$ | ~80 | -0.16 | 10 | -0.016 | [10] |
| $PZO^c$ | 211 | -3.5 | 28 | -0.125 | **This work** |
| Mn-doped $PZT^c$ | - | 0.55 | 60 | 0.009 | [11] |
| $PZNST^c$ | 47 | 1.1 | 50 | 0.022 | [7] |
| $PST35^c$ | 323 | 2.05 | 40 | 0.051 | [12] |
| $BZT^c$ | 113 | 2.4 | 30 | 0.080 | [13] |
| 0.5BZT-$0.5BCT^c$ | 100 | 0.46 | 60 | 0.008 | [14] |
| $NBT^m$ | 90 | 1.7 | 90 | 0.019 | [15] |
| $BNT^c$ | 40 | 1.04 | 50 | 0.021 | [16] |
| Sn-doped $BST^c$ | 67 | 0.19 | 7 | 0.027 | [17] |
| $BSTZS^c$ | 30 | 0.22 | 7.5 | 0.029 | [18] |
| Ba-doped $SBNT^c$ | 140 | 0.78 | 50 | 0.016 | [19] |
| $SBN75^s$ | 80 | 0.4 | 10 | 0.040 | [20] |
| $BTO^c$ | 120 | 0.14 | 3 | 0.047 | [21] |
| $BTO^m$ | 97 | 0.94 | 308 | 0.003 | [22] |
| $BTO^s$ | 140 | 1.6 | 10 | 0.160 | [23] |
| La-doped PMN-$12PT^c$ | 110 | 2.6 | 30 | 0.087 | [24] |
| PMN-$10PT^c$ | 25 | 0.23 | 100.5 | 0.002 | [25] |
| PMN-$30PT^s$ | 135 | 0.6 | 10 | 0.060 | [18] |
| PMN-$30PT^s$ | ~137 | 0.65 | 10 | 0.065 | [10] |
| $PST50^m$ | 57 | 5.5 | 290 | 0.019 | [26] |
| $PZO^c$ | 226 | 5.6 | 42 | 0.133 | **This work** |

*Table S1 Compilation of negative and positive electrocaloric effects measured by direct methods. Superscripts c, m and s corresponds to ceramics, multilayer capacitors and single crystals respectively.*

Finally, figure S7 shows the ON/OFF electrocaloric asymmetry in the region where the negative electrocaloric effect is large, and also the different mechanisms in the different regions corresponding to figure 1-b in the manuscript.

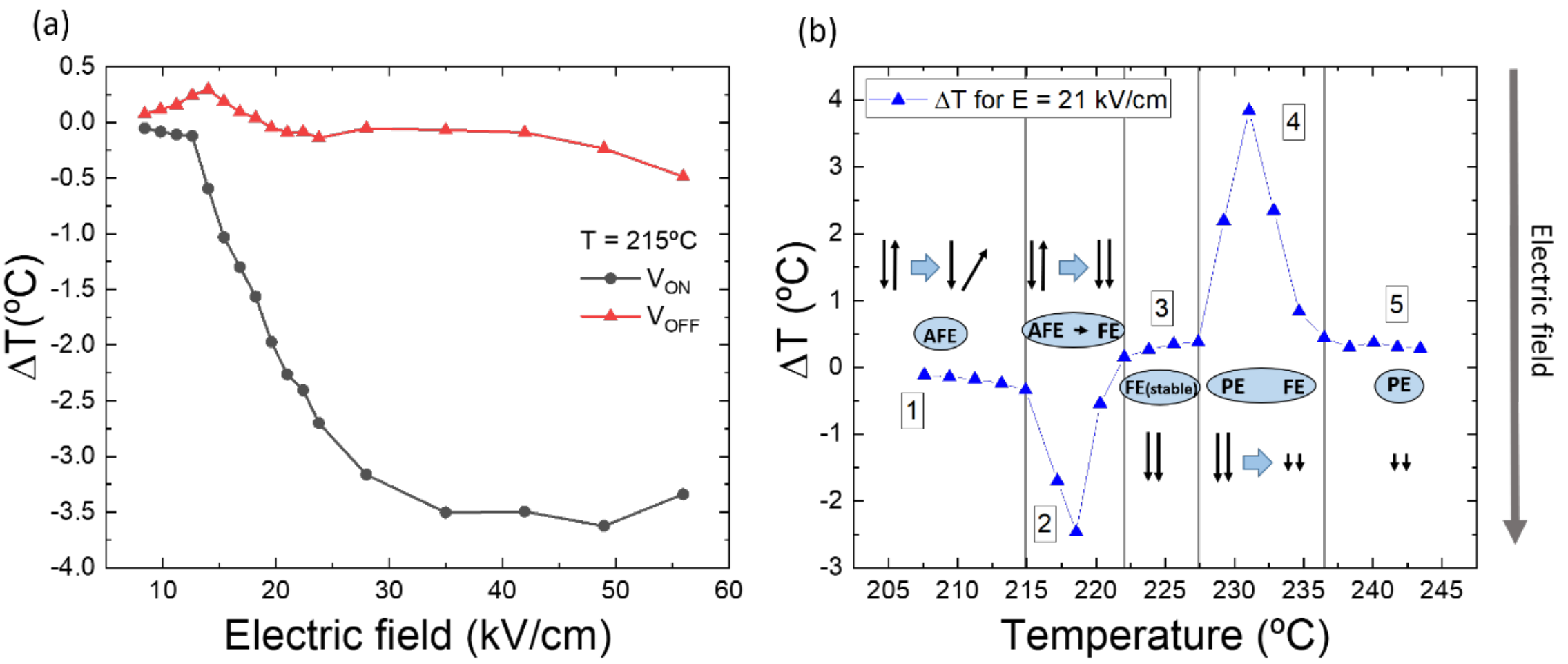

*Figure S7 (a) Electrocaloric temperature change versus applied electric field at constant temperature. The large asymmetry between the on/off state reflects the phase transition nature of the process. (b) Schematic representation of the different mechanisms in each region of the electrocaloric response.*

## SECTION S2: COMPLETE DYNAMIC MECHANICAL ANALYZER (DMA) DATA

In the following figures both heating and cooling curves of the storage modulus of $PbZrO_3$ as a function of applied electric field and sample's temperature will be displayed. The details on the experimental procedure are explained in the Materials and Methods section in the manuscript.

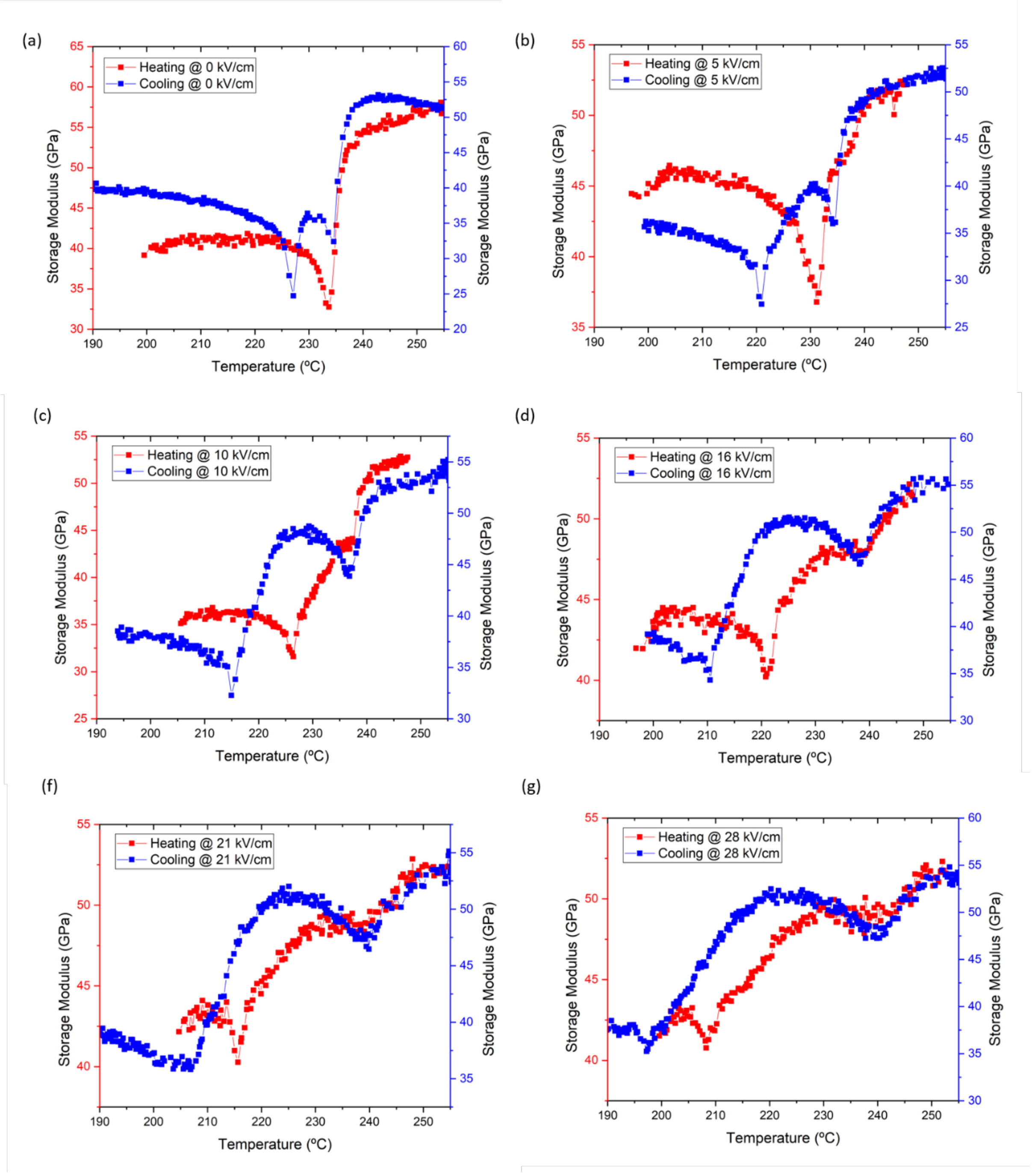

*Figure S8 Storage modulus of $PbZrO_3$ ceramics as a function of temperature on both heating and cooling processes for 6 different electric fields.*

## SECTION S3: RAW THERMAL DATA

Infrared raw data is shown in **figure S8** with different electrocaloric responses along the phase diagram of PbZrO3 ceramics. To summarize in just one field all the possible scenarios, we picked the curve for 35 kV/cm, as for this field we have both a low and high negative ECE (reaching ΔT values greater than -3K), followed by a low and high positive negative ECE. In Figure S9(a) we show the Joule heating at 28 kV/cm for both negative and positive effects where the leakage is larger. Note that, as specified in the manuscript the large response of the negative and positive electrocaloric effect are repeatable, but asymmetric in their ON/OFF cycle, as shown in **Figure S8-c,f) and Figure S9-b**.

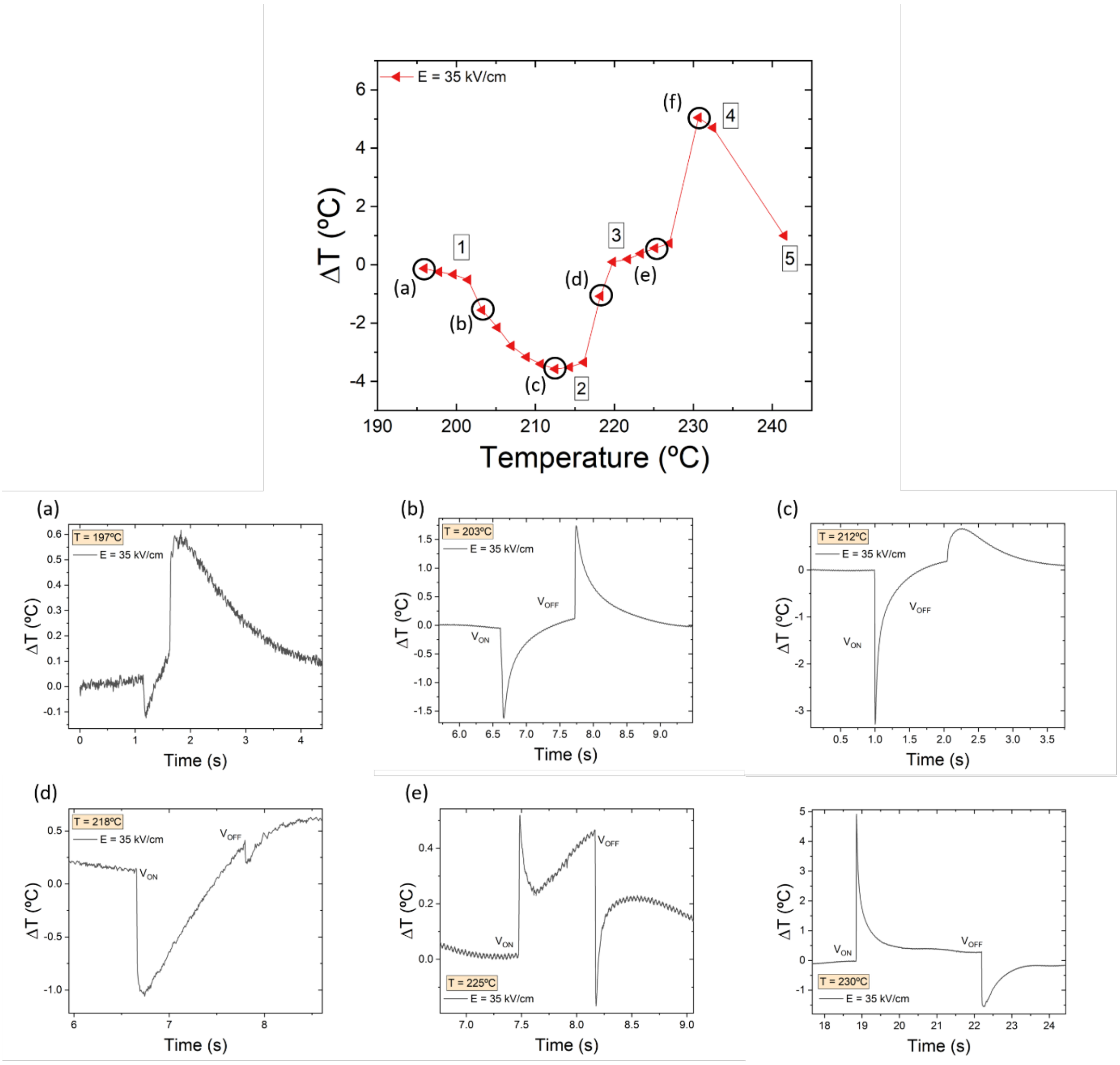

*Figure S9 Raw data for different points in the electrocaloric data of $PbZrO_3$ ceramics for an electric field of 35 kV/cm: (a)-(d) correspond to different magnitudes of the negative ECE while (e)-(f) to the positive ECE.*

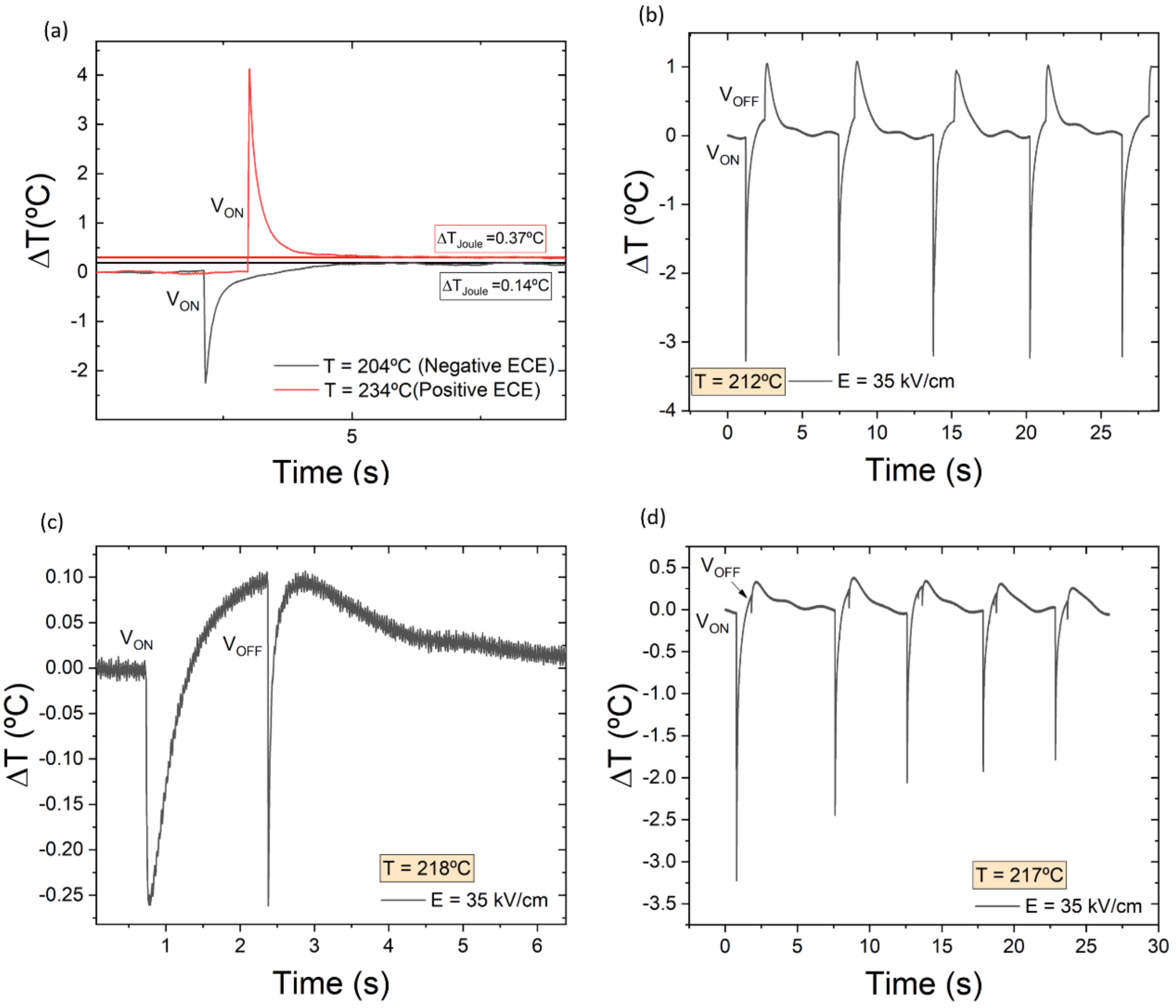

*Figure S9 (a) Joule heating generated in the negative and positive electrocaloric regime for an electric field E = 24 kV/cm; (b) sequential On-Off Brayton cycles of the negative electrocaloric effect showing repeatability in ΔT, (c) On-Off sequence showing an anomalous Off response conformed by a fast ΔT<0 and (d) On-Off cycles showing a decrease in the electrocaloric response due to the creation of stable FE domains that decrease the overall latent heat.*

However, in the transition from region 2 to 3 (region (d) in Figure S8) this repeatable cycles are not anymore (Figure S9-d). This is directly related to the irreversible transition of domains from their ground AFE phase to a stable (or metastable) FE phase.

This means that, once the voltage is turned ON for the first time, some domains in the sample will stay in their FE phase without going back to an AFE structure; this fact has two consequences:

(i) The next time the voltage is turned ON there will be less AFE domains to be switched, so the latent heat released in the AFE-FE phase transition will be less, and the ECE response will decrease with respect to the previous cycle **(see $V_{ON}$ peaks in Figure S9(d))**.

(ii) When the voltage is turned OFF, the electrocaloric response will be that of a FE phase and thus, it will yield a $\Delta T < 0$ (Figure S8-d and Figure S9-c)

## SECTION S4: DIFFERENTIAL SCANNING CALORIMETRY (DSC) DATA

From the heat flux measurements in figure 4 in the manuscript, we can construct a similar phase diagram as already done with mechanical and Infrared data in Figure 5-a. In figure S10 we show the Temperature-Electric field diagram with the AFE-FE and FE-PE phase transitions on heating, which are in agreement with the DMA data.

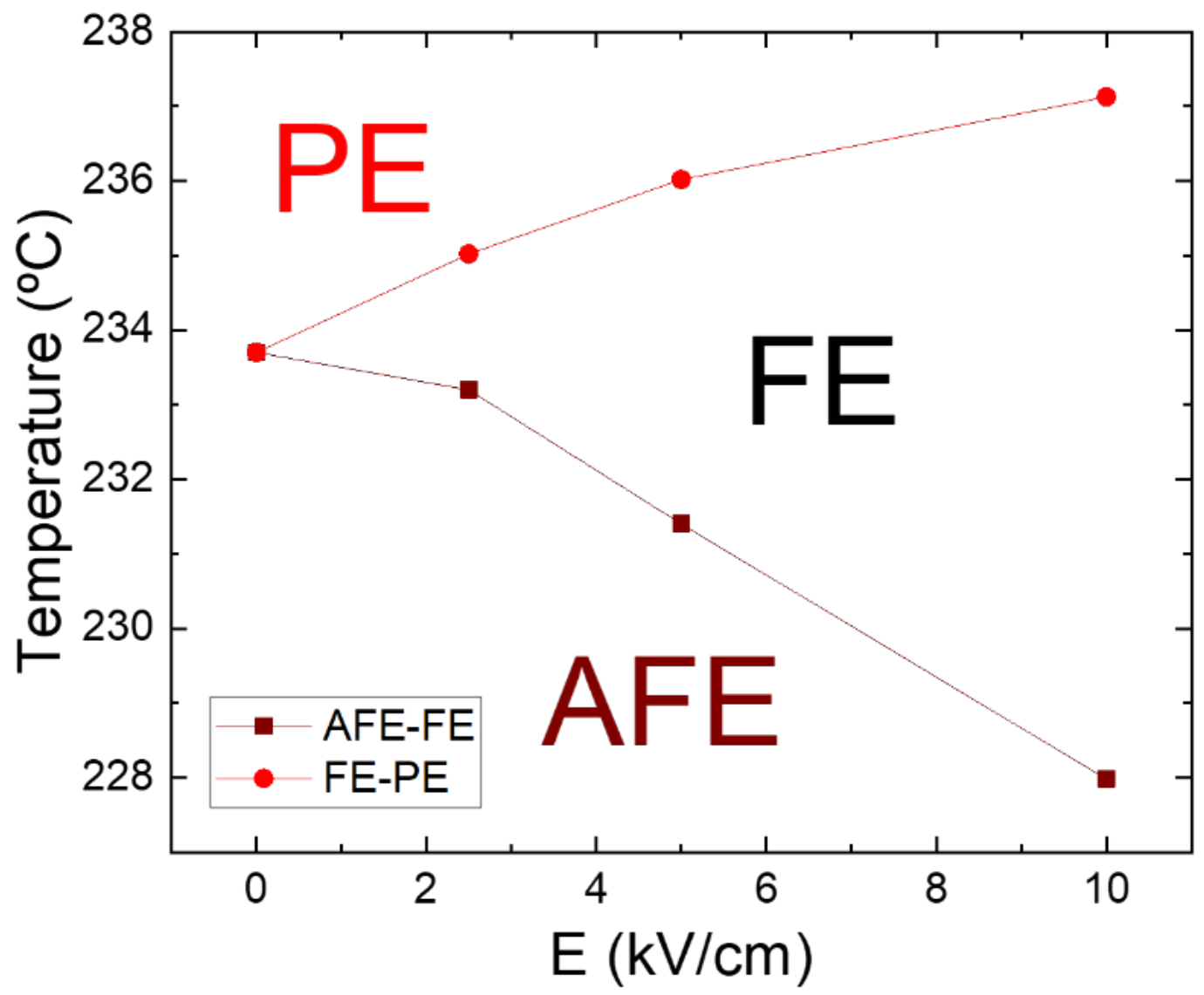

*Figure S10 Phase diagram constructed with DSC data. The different phase regions are labeled as AFE (antiferroelectric), ferroelectric (FE) and paraelectric (PE)*